\documentclass[fleqn,usenatbib]{mnras}

\usepackage[T1]{fontenc}
\usepackage{ae,aecompl}

\usepackage{graphicx}	
\usepackage{amsmath}	
\usepackage{amssymb}	
\usepackage{float}
\usepackage{fancyhdr}
\usepackage{hyperref}
\usepackage{longtable}
\usepackage{subfig}
\usepackage{graphicx}
\usepackage[usenames,dvipsnames]{xcolor}

\usepackage{soul}
\newcommand{\del}[1]{\textcolor{gray}{\st{#1}}} 

\usepackage{epstopdf}

\newcommand{\Teff}{\mbox{$T_{\mathrm{eff}}$}}

\newcommand{\hel}[2] {He\,{\sc #1}~$\lambda$#2}
\newcommand{\Line}[3]{#1\,{\sc #2}~$\lambda$#3}

\newcommand{\molec}[2]{#1$_{\sc #2}$}
\newcommand{\Ion}[2]{#1\,{\sc #2}}

\newcommand{\kms}{\mbox{$\mathrm{km\,s^{-1}}$}}

\newcommand{\Msun}{\mbox{$\mathrm{M}_{\odot}$}}

\newcounter{tref}

\newcommand{\WD}{WD\,1145+017}
\newcommand{\wds}{WD1145}
\newcommand{\degree}{$^\mathrm{o}$}

\title[Fast spectrophotometry of \WD]{Fast spectrophotometry of WD\,1145+017 }

\author[Paula Izquierdo et al.]{Paula Izquierdo$^{1,2}$\thanks{E-mail: pizdo@iac.es},
Pablo Rodr\'iguez-Gil$^{1,2}$,
Boris T. G\"ansicke$^{3,4}$, 
Alexander J. Mustill$^{5}$,
\newauthor
Odette Toloza$^{3}$,
Pier-Emmanuel Tremblay$^{3}$,
Mark Wyatt$^{6}$,
Paul Chote$^{3}$,
Siegfried Eggl$^{7}$,
\newauthor
Jay Farihi$^{8}$,
Detlev Koester$^{9}$,
Wladimir Lyra$^{7,10}$,
Christopher J. Manser$^{3}$,
\newauthor
Thomas R. Marsh$^{3}$,
Enric Pall\'e$^{1,2}$,
Roberto Raddi$^{11}$,
Dimitri Veras$^{3,4}$\thanks{STFC Ernest Rutherford Fellow},
\newauthor
Eva Villaver$^{12}$,
Simon Portegies Zwart$^{13}$
\\
$^{1}$Instituto de Astrof\'isica de Canarias, 38205 La Laguna, Tenerife, Spain\\
$^{2}$Departamento de Astrof\'isica, Universidad de La Laguna, 38206 La Laguna, Tenerife, Spain\\
$^{3}$Department of Physics, University of Warwick, Coventry CV4 7AL, UK\\
$^{4}$Center for Exoplanets and Habitability, University of Warwick, Coventry CV4 7AL, UK\\
$^{5}$Lund Observatory, Department of Astronomy $\&$ Theoretical Physics, Lund University, Box 43, SE-221 00 Lund, Sweden\\
$^{6}$Institute of Astronomy, Madingley Rd, Cambridge CB3 0HA, UK\\
$^{7}$Jet Propulsion Laboratory, California Institute of Technology, 4800 Oak Grove Drive, 91109 Pasadena, CA, USA\\
$^{8}$Physics and Astronomy, University College London, London, WC1E 6BT, UK\\
$^{9}$Institut f\"ur Theoretische Physik und Astrophysik, Universit\"at Kiel, 24098, Kiel, Germany\\
$^{10}$California State University, Northridge, Department of Physics and Astronomy, 18111 Nordhoff St, Northridge, CA, 91330\\
$^{11}$Dr. Karl Remeis-Sternwarte, Friedrich Alexander Universit\"at Erlangen-N\"urnberg, Sternwartstr. 7, 96049 Bamberg, Germany\\
$^{12}$Departamento de F\'\i sica Te\'orica, Universidad Aut\'onoma de Madrid, Cantoblanco 28049 Madrid, Spain\\
$^{13}$Sterrewacht Leiden, Leiden University, P.O. Box 9513, 2300 RA Leiden, The Netherlands
}

\date{Accepted XXX. Received YYY; in original form ZZZ}

\pubyear{2018}

\begin{document}
\label{firstpage}
\pagerange{\pageref{firstpage}--\pageref{lastpage}}
\maketitle

\begin{abstract}
WD\,1145+017 is currently the only white dwarf known to exhibit periodic transits of planetary debris as well as absorption lines from circumstellar gas. We present the first simultaneous fast optical spectrophotometry and broad-band photometry of the system, obtained with the Gran Telescopio Canarias (GTC) and the Liverpool Telescope (LT), respectively. The observations spanned $5.5$\,h, somewhat longer than the $4.5$-h orbital period of the debris. Dividing the GTC  spectrophotometry into five wavelength bands reveals no significant colour differences, confirming grey transits in the optical. We argue that absorption by an optically thick structure is a plausible alternative explanation for the achromatic nature of the transits that can allow the presence of small-sized ($\sim\mu$m) particles. The longest ($87$\,min) and deepest ($50$\,per\,cent attenuation) transit recorded in our data exhibits a complex structure around minimum light that can be well modelled by multiple overlapping dust clouds. The strongest circumstellar  absorption line, \Line{Fe}{ii}{5169}, significantly weakens during this transit, with its equivalent width reducing from a mean out-of-transit value of $2$\,\AA\ to $1$\,\AA\ in-transit, supporting spatial correlation between the circumstellar gas and dust. Finally, we made use of the \textit{Gaia} Data Release~2 and archival photometry to determine the white dwarf parameters. Adopting a helium-dominated atmosphere containing traces of hydrogen and metals, and a reddening $E(B-V)=0.01$ we find $T_\mathrm{eff}=15\,020 \pm 520$\,K, $\log g=8.07\pm0.07$, corresponding to $M_\mathrm{WD}=0.63\pm0.05\,\Msun$ and a cooling age of $224\pm30$\,Myr. 
\end{abstract}





\section{Introduction}
\label{intro}
About $25-50$ per cent of all white dwarfs contain traces of elements heavier than helium in their atmospheres \citep{Zuckerman03,zuckerman10-1,Koester14}. Because of the strong surface gravity of white dwarfs, metals will sink out of their atmospheres on time scales that are short compared to their cooling ages \citep{paquetteetal86-1,koester09-1}. This characteristic implies that the photospheric pollution is the result of recent or ongoing accretion, and it is now well-established that the accreted material originates from the disruption of planetary bodies \citep{graham90-1,jura03}. A small fraction of white dwarfs, $\simeq1-3$ per cent, show along with the metal-pollution an infrared excess produced by a circumstellar dust disc \citep[see e.g.][]{kilic06, girvenetal11-1, steeleetal11-1}, which unambiguously identifies the source of the metals detected in their photospheres. 


The discovery of periodic transits from debris orbiting the white dwarf \WD~(hereafter \wds; \citealt{vanderburg15}) provides the unprecedented opportunity to investigate the physical processes of the tidal disruption of a planetesimal in real time. Detailed follow-up photometry revealed the presence of multiple debris fragments with periods of $4.5-4.9$\,h, many of them co-orbital with period differences of only a few seconds \citep{gansicke16-1,rappaportetal16-1,garyetal17-1}. The variable and large depths (up to 50 per cent) of the occultations, coupled with their long durations (a few to a few tens of minutes) and asymmetric shapes, suggest that the transits are caused by dust clouds originating from multiple fragments rather than the solid bodies themselves \citep{gansicke16-1}. It has been suggested that a single parent body may be the origin of the multiple fragments \citep{rappaportetal16-1} and upper limits on the mass ($\le10^{23}$\,g) and eccentricity ($\le0.1$) of the putative body were derived using the dynamical information obtained from the timing of the transits \citep{gurrietal17-1,verasetal17-2}.

High-resolution spectroscopy of \wds~revealed a metal-polluted He-dominated atmosphere, with signatures from 12 elements: \Ion{O}{}, \Ion{Mg}{}, \Ion{Al}{}, \Ion{Si}{}, \Ion{Ca}{}, \Ion{Ti}{}, \Ion{V}{}, \Ion{Cr}{}, \Ion{Mn}{}, \Ion{Fe}{}, \Ion{Ni}{} and \Ion{Na}{} \citep{xu16-1}. Besides the presence of photospheric metal lines, they detected numerous absorption features from circumstellar gas. \citet{redfieldetal17-1} reported a change in the shape of these circumstellar absorption lines on time scales ranging from minutes to months, and tentatively associated this evolution with an eccentric gas disc, a hypothesis that was more recently refined with additional observations and modelling \citep{cauleyetal18-1}.


Numerous simultaneous multi-band photometric studies of \wds\ revealed no measurable difference in the transit depths across different band passes \citep{alonso16-1,zhouetal16-1,crolletal17-1,xuetal18-1}, which provided lower limits on the particle size in the range of $\simeq0.5-1\,\mu\mathrm{m}$. However, \citet{hallakounetal17-1} reported a \textit{decrease} in the transit depth in the $u$-band, and suggested that this `bluing' is related to a reduced strength of the numerous circumstellar absorption lines present in the very blue part of the optical spectrum. The previously reported apparent decrease in the strength of the circumstellar Fe absorption lines during transit events \citep{redfieldetal17-1} provides some support for this hypothesis, though there was a lack of simultaneous photometry to confirm this association.

Here we report the first simultaneous spectroscopic and photometric study of \wds, obtaining fast optical spectrophotometry and photometry using two different telescopes.



\section{Observations and Data Reduction}

We observed \wds\ on 2017 April 4 simultaneously with the $10.4$-m Gran Telescopio Canarias (GTC) and the $2.0$-m Liverpool Telescope (LT), both located at the Observatorio del Roque de los Muchachos on La Palma, Spain. The observations were conducted under clear sky conditions, with variable seeing between $1.1$ and $1.7\ \mathrm{arcsec}$.

\subsection{GTC spectrophotometry}

For the differential spectrophotometry we used the Optical System for Imaging and low-Intermediate-Resolution Integrated Spectroscopy \citep[OSIRIS;][]{10.1117/12.926676}, mounted on the Nasmyth-B focus of GTC. We used the $12$-$\mathrm{arcsec}$ wide slit in order to minimise flux losses, and the R300B grism. The spectra were recorded on CCD \#2 (Marconi CCD$44$--$82$, $2048 \times 4096$ pixels) with $2\times2$ binning. This setup provides a spectral resolution of 21\,\AA\ (full-width at half maximum; FWHM) at the centre of the $3965-9575$\,\AA\ wavelength range.

We orientated the slit at $61.7$\degree\ (east from north) to include both \wds\ and the reference star UCAC4 458--051099 (1.9 arcmin distant). We took $250$ exposures of 50\,s each with a dead time of 21\,s, for a total on-source time of 5.16\,h. The observing run started at airmass $1.38$ and ended at 1.55, before and after target culmination, respectively. We tested that no significant flux drop was present for both stars. 

\begin{figure*}
	\includegraphics[width=2.12\columnwidth]{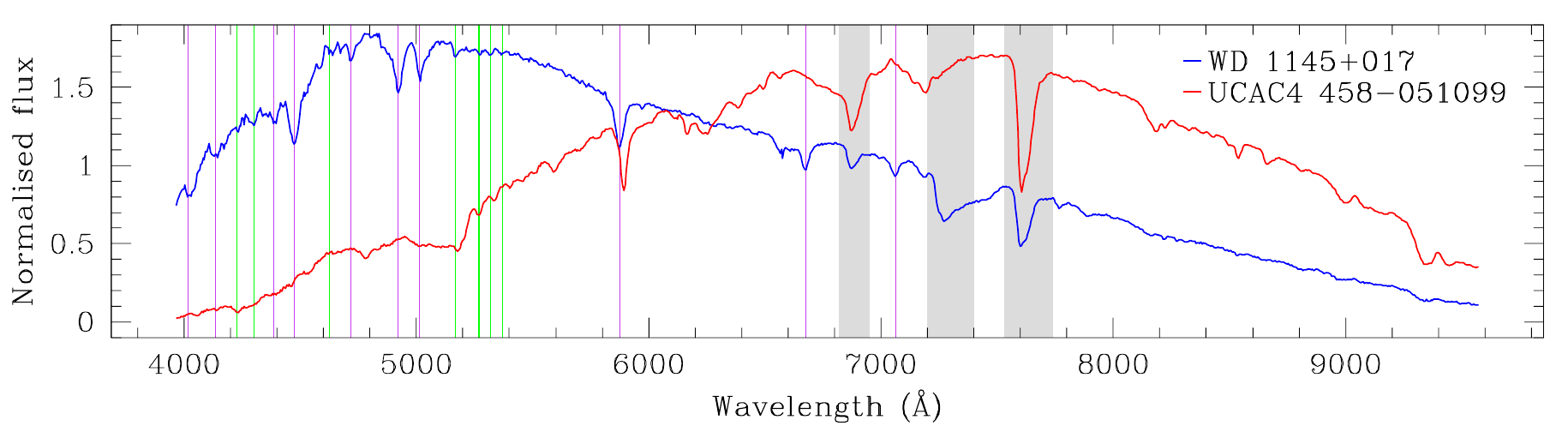}
    \caption{Average wide-slit spectra of \wds\ (blue) and the reference star (red) obtained with GTC/OSIRIS. The \Ion{He}{i} and \Ion{Fe}{ii} absorption features of \wds\ are marked with purple and green vertical lines, respectively. The principal telluric absorptions lie within the light-grey shaded areas. Both average spectra have been normalised. }
    \label{fig:spectra}
\end{figure*}

The spectra were bias and flat-field corrected using standard procedures within {\sc iraf}\footnote{{\sc iraf} is distributed by the National Optical Astronomy Observatories.}. We obtained the one-dimensional spectra of both objects via optimal extraction \citep{horne86-1} using the data reduction software \del{PAMELA} {\sc pamela}\footnotemark. We adopted an aperture of 32 pixels, centred on each spectrum trace. Two $10$-pixel wide apertures bracketing each target aperture were defined to subtract the sky background. Wavelength calibration of the spectra was performed with {\sc molly}\footnotemark[\value{footnote}] by means of a fourth-order polynomial fit to the HgArNeXe arc data. We corrected  for any significant wavelength shifts by applying the offsets measured on the prominent \hel{i}{5876} for \wds\ and the \Line{Na}{i}{5890,\,5896} (blended) absorption doublet for the reference star. The average spectra for the target and the reference star are shown in Fig.\,\ref{fig:spectra}.

\footnotetext{Both {\sc pamela} and {\sc molly} packages were developed by Tom Marsh; \url{http://deneb.astro.warwick.ac.uk/phsaap/software/molly/html/INDEX.html}}



\subsection{LT fast photometry}

We obtained simultaneous fast photometry of \wds\ with RISE \citep{gibsonetal08-1,10.1117/12.787889}, a fast-readout camera on the LT. RISE was equipped with the `$V$+$R$' filter which has a bandpass of $5000-7000$\,\AA and a $1024\times 1024$ pixel e2v frame-transfer CCD, which provides a field of view of $9.4\times9.4\ \mathrm{arcmin}^{2}$. The frame-transfer operation rapidly shifts the integrated charge into a buffer region at the end of each exposure, where it can be read out in parallel with the next integration. We stayed on-target for a total of 5.03\,h, continuously taking 5\,s exposures with a negligible 0.035\,s dead time between images. Debiasing and flat-fielding were performed by the RISE data reduction pipeline, then synthetic aperture photometry was extracted for the target and several nearby comparison stars using the {\sc tsreduce} software \citep{chote14}.  A 2-pixel aperture radius was chosen to minimise noise in the target over (sum of) comparison flux-ratio. 





\section{Results}


\subsection{Atmospheric parameters of \wds}
Spectroscopic determination of the surface gravity ($\log g$) of helium-atmosphere white dwarfs is subject to systematic uncertainties (e.g. \citealt{koester+kepler15-1}), and the published studies of \wds\ adopted a canonical $\log g=8$ \citep{vanderburg15,xu16-1}. The Data Release~2 (DR2) of the \textit{Gaia} mission \citep{gaiaDR2-1} provides an accurate parallax of \wds, $\pi=7.06\pm0.12$\,mas, placing it at a distance\footnote{The parallax is sufficiently precise that the distance can simply be computed from its inverse \citep{bailer-jonesetal18-1}.} of $d=141.7 \pm 2.5$\,pc. We have estimated the atmospheric parameters using the \textit{Gaia} parallax and three independent photometric data sets: \textit{Gaia} $G$, $G_{\rm BP}$, $G_{\rm RP}$, Pan--STARRS $g$, $r$, $i$, $z$, $y$, and SDSS $u$, $g$, $r$, $i$, $z$ (Table\,\ref{tab:phot}). For this purpose we computed a grid of synthetic DBA white dwarf atmosphere models, adopting a helium-dominated composition containing some trace hydrogen ($\mathrm{\log \,[H/He]}=\,-4.70$, \citealt{xu16-1}). The grid spanned effective temperatures of $\Teff=13\,000-17\,000$\,K in steps of 250\,K and surface gravities of $\log{g}=7.5-8.5$ in steps of 0.1\,dex. To explore the effect of metal contamination on the derived atmospheric parameters, we also computed a grid of DBAZ models with the same range in \Teff\ and $\log g$, but including trace-metals with abundances as measured by \citet{xu16-1}.

The fitting technique was the same as  described in \citet{hollands18}. The synthetic fluxes were thereafter combined with the mass-radius relation of \citet{fontaine01} for thin hydrogen layers ($M_{\rm H}/M_{\rm WD}$ = 10$^{-10}$, with $M_\mathrm{H}$ and $M_\mathrm{WD}$ the masses of the hydrogen layer and the white dwarf, respectively) to predict the stellar flux. This fixed the stellar radius $R_{\mathrm{WD}}$ for a given $T_{\rm eff}$ and $\log{g}$. We then minimised the $\chi^2$ between absolute {\it Gaia} photometry and absolute synthetic magnitudes to derive the atmospheric parameters for each data set. Given that these fits to photometric data are sensitive to interstellar extinction, we repeated the fits for two adopted values of reddening, the total column along the line-of-sight towards \wds\ \citep{schlafly+finkbeiner11-1}, and the most likely column given the distance to \wds\ based on the three-dimensional reddening maps of \citet{greenetal15-1} and \citet{capitanioetal17-1}. Cooling ages were computed using an updated set of models described by \citet{fontaine01}. Table\,\ref{tab:params} summarises the best-fit parameters, and illustrates that all three photometric data sets result in parameters that are consistent with each other within the statistical errors, and that the main uncertainty is the so far poorly constrained amount of interstellar extinction. 

We adopt the mean of the parameters for the most likely reddening, $E(B-V)=0.01$, obtaining: $T_\mathrm{eff}=15020 \pm 520\,(15380 \pm 370)$\,K and $\log g=8.07\pm0.07\,(8.12\pm0.05)$ for the DBAZ\,(DBA) atmosphere model. Using the cooling models of \citet{fontaine01}, these parameters correspond to $M_\mathrm{WD}=0.63\pm0.05\,(0.66\pm0.03)\,\Msun$ and a cooling age of $224\,\pm30\,(228\pm20)$\,Myr. Including photospheric metals in the fit implies significant additional opacities in the ultraviolet wavelength range, resulting in strong line blanketing compared to the DBA models, and consequently in a slightly lower effective temperature. Given the fixed distance, this lower temperature is compensated by a slightly larger radius, and hence lower mass. A final assessment of the white dwarf parameters will have to await the analysis of ultraviolet spectroscopy of \wds\ to account in detail for the effect of metal-line blanketing. 

The absorption lines from the circumstellar gas \citep{xu16-1} will cause minor perturbations in the observed photometric spectral energy distribution of \wds. The results derived from the fits to three independent sets of photometry obtained over 15 years (SDSS: 2001, Pan--STARRS: 2012, Gaia: 2015) are consistent within their statistical errors, demonstrating that the effect of the reported variability in the strength of the circumstellar lines \citep{redfieldetal17-1, cauleyetal18-1} is irrelevant for the atmospheric parameters compared to the other sources of uncertainty.


We note that the cooling age of \wds\ places it in a phase of the post-main sequence evolution where enhanced dynamical activity is expected in the population of planetesimals around white dwarfs \citep{mustilletal18-1}. Using the initial-to-final mass relations of \citet{catalanetal08-2}, \citet{kaliraietal08-1}, \citet{casewelletal09-1} and \citet{williamsetal09-1}, these white dwarf parameters imply a progenitor of $M_\mathrm{MS}=2.48\pm0.14\,\mathrm{M}_\odot$, where $M_\mathrm{MS}$ was the mass of the star in the main sequence. This corresponds to an early A-type star with a main-sequence lifetime of $550\pm100$\,Myr, hence the total age of the system is a bit below one Gyr.

\begin{table}
%
\begin{tabular}{ccc}
\hline
\hline
\textit{Gaia} DR2 & Pan--STARRS & SDSS \\
\hline\noalign{\smallskip}
$G_\mathrm{BP} = 17.039 \pm 0.012$        &                          & $u=16.932 \pm 0.028$   \\
$G_\mathrm{~~\,\,} = 17.167 \pm 0.002$    & $g = 17.072 \pm 0.001 $  & $g=17.076 \pm 0.016$   \\
$G_\mathrm{RP} = 17.234 \pm 0.010$        & $r = 17.321 \pm 0.007 $  & $r=17.377 \pm 0.017$   \\
                                          & $i = 17.563 \pm 0.007$   & $i=17.601 \pm 0.013$   \\
                                          & $z = 17.825 \pm 0.009$   & $z=17.838 \pm 0.024$   \\
                                          & $y = 17.955 \pm 0.022$   &                        \\
\noalign{\smallskip}\hline
\end{tabular}
\caption{Photometric magnitudes used to derive the atmospheric parameters of \wds. The Pan--STARRS and SDSS data are given in mean-PSF (Point Spread Function; \citet{PanSTARRS2}) and PSF mag \citep{SDSSpass}, respectively.}
\label{tab:phot}
\end{table}


\begin{table*}
\centering
\begin{tabular}{cccccc}
\hline
\hline
Parameter            & Model               & {\it Gaia} DR2                     & Pan--STARRS                          & SDSS                                & Mean value\\
                     &                     & $E(B-V)=0.01/0.024$                & $E(B-V)=0.01/0.024$                 & $E(B-V)=0.01/0.024$                 & $E(B-V)=0.01/0.024$\\
\hline
$T_\mathrm{eff}$ (K) & DBA                 & $15440 \pm 770$/$15870 \pm 840$    & $15300 \pm 760$/$15890 \pm 890$     &	$15400 \pm 240$/$16040 \pm 270$     & $15380 \pm 370$/$15940 \pm 420$\\
                     & DBAZ                & $15090 \pm 1050$/$15640 \pm 1040$  & $14890 \pm 1110$/$15640 \pm 1090$   &	$15070 \pm 330$/$15840 \pm 320$     & $15020 \pm 520$/$15710 \pm 510$\\
\hline
$\log g$        & DBA                      & $8.12 \pm 0.10$/$8.14 \pm	0.11$   & $8.13	\pm	0.10$/$8.16 \pm	0.11$     &	$8.12 \pm 0.03$/$8.16 \pm 0.02$ & $8.12 \pm 0.05$/$8.15 \pm 0.05$ \\
                & DBAZ                     & $8.07 \pm 0.15$/$8.11 \pm	0.14$   & $8.07	\pm	0.16$/$8.13	\pm	0.14$     &	$8.07 \pm 0.02$/$8.14 \pm 0.03$ & $8.07 \pm 0.07$/$8.13 \pm 0.07$ \\
\hline
$M_\mathrm{WD}$ (M$_{\odot}$)  & DBA       & $0.66 \pm 0.06$/$0.68 \pm 0.07$   & $0.67 \pm 0.06$/$0.69 \pm 0.07$  & $0.662	\pm	0.016$/$0.689 \pm 0.016$ & $0.66 \pm 0.03$/$0.69 \pm 0.03$\\
                               & DBAZ      & $0.63 \pm 0.09$/$0.66 \pm 0.08$   & $0.63 \pm 0.10$/$0.67 \pm 0.08$   & $0.634	\pm	0.016$/$0.673 \pm 0.016$ & $0.63 \pm 0.05$/$0.67 \pm 0.04$  \\
\hline
$R_\mathrm{WD}$ ($0.01$\,R$_{\odot}$) & DBA   & $1.18 \pm 0.08$/$1.16 \pm 0.08$ & $1.17 \pm 0.08$/$1.14 \pm 0.08$  & $1.17 \pm 0.02$/$1.14	\pm	0.02$    & $1.17 \pm 0.04$/$1.15 \pm 0.04$ \\
                                      & DBAZ  & $1.21 \pm 0.12$/$1.18 \pm 0.11$ & $1.21 \pm 0.13$/$1.16 \pm 0.11$  & $1.21 \pm 0.02$/$1.16	\pm	0.02$    &$1.21 \pm 0.06$/$1.17 \pm 0.05$ \\
\hline
Age (Myr)                       & DBA      & $223 \pm 41$/$214 \pm 42$      & $235 \pm 42$/$223 \pm 43$     & $227 \pm 11$/$216	\pm	10$ & $228 \pm 20$/$218 \pm 20$\\
                                & DBAZ     & $220 \pm 59$/$213 \pm 53$      & $230 \pm 65$/$222 \pm 55$     & $223 \pm 10$/$215	\pm	10$ & $224 \pm 30$/$217 \pm 26$\\

\hline
\end{tabular}
\caption{Atmospheric parameters of \wds~derived from fitting the \textit{Gaia}~DR2 parallax and the photometry from \textit{Gaia}~DR2, Pan--STARRS, and SDSS, adopting two values for the extinction, the total column ($E(B-V)=0.024$, \citealt{schlafly+finkbeiner11-1}), and the likely column for $d=141$\,pc based on three-dimensional reddening maps ($E(B-V)=0.01$, \citealt{greenetal15-1,capitanioetal17-1}). Parameters were derived from helium-dominated atmosphere models without (DBA) and with metals (DBAZ), with the abundances fixed to those determined by \citet{xu16-1}.}
\label{tab:params}
\end{table*}

\subsection{Simultaneous GTC and LT `white-light' light curves}
\label{sec:whitelcs}

\begin{figure*}
	\includegraphics[width=2.12\columnwidth]{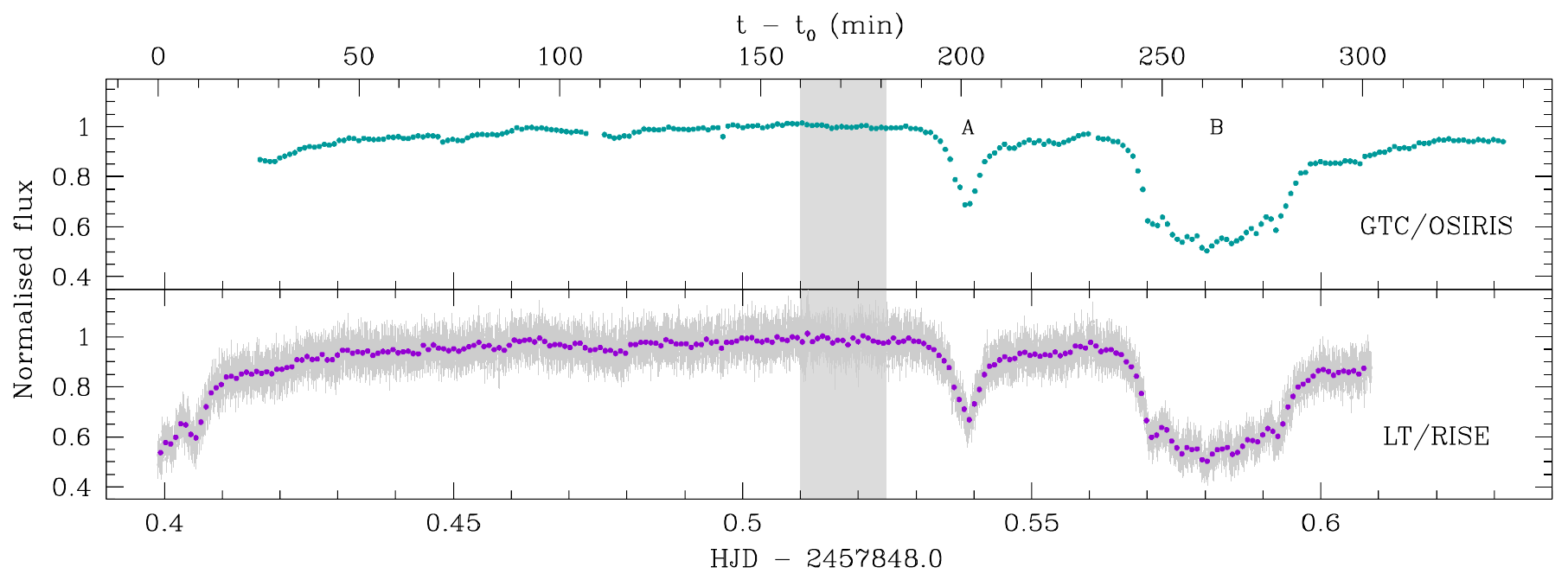}
    \caption{GTC/OSIRIS (top) and LT/RISE (bottom) light curves obtained on 2017 April 4. The GTC light curve was obtained by integrating the spectra over $4300$--$9200$\,\AA. The dark purple dots in the LT light curve were binned by a factor 15 to obtain a cadence similar to that of the GTC observations. The top $x$-axis ($t-t_{0}$) is the time in minutes from the first data point of the unbinned LT light curve. The median of the out-of-transit points inside the grey shaded area has been used to normalise the light curves. The two transits observed are labelled as `A' and `B'.}
    \label{fig:lt_gtc_LCs}
\end{figure*}

Every light curve in this paper obtained from the GTC data has been derived by integrating the flux over a limited wavelength range for each individual spectrum. We avoided the three telluric regions marked by light-grey shaded areas in Fig.\,\ref{fig:spectra} for the flux sum: two produced by \molec{O}{2} ($6820-6950$ and $7530-7740$\,\AA), and a water vapour feature ($7200-7400$\,\AA). We then computed the integrated flux ratio between \wds~and the reference star. Finally, we normalised these flux-ratio curves to the median value of the out-of-transit points that lie in the interval between 159.9 and 181.5\,min, shown as a light-grey shaded area in the light curve plots.


The `white-light' GTC light curve was obtained by integrating the flux over the wavelength range $4300-9200$\,\AA. Figure\,\ref{fig:lt_gtc_LCs} shows the GTC spectrophotometry along with the LT light curve, which we binned by a factor of 15, resulting in a cadence similar to the GTC observations.

\begin{table}
  \centering
\begin{tabular}{ccccc}
\hline
\hline
Dip & Telescope/ & Start -- End  & $\Delta t$  & Depth  \\
  & Instrument & (HJD -- 2457848.0) & (min) & ($\%$) \\
\hline
A & GTC/OSIRIS  & 0.5333 -- 0.5572 & 34.5 & 31.2 \\
   & LT/RISE     & 0.5322 -- 0.5575 & 36.4  & 30.3  \\
B & GTC/OSIRIS  & 0.5615 -- 0.6221 & 87.3 & 49.7  \\
   & LT/RISE     & 0.5611 --  &  - & 49.5  \\
\hline
\end{tabular}
\caption{Transit (dip) identification, telescope/instrument used, start and end times, total time duration ($\Delta t$) and relative depth at minimum light relative to the out-of-transit level. The normalised flux threshold defining the transit start and end points was taken as $0.95$.}
\label{tab:dips}
\end{table}

Both light curves show consistent behaviour. They exhibit two  distinct transit events (labelled as `A' and `B' in the figure), whose start and end times as well as depths are listed in Table\,\ref{tab:dips}. The beginning of the LT light curve displays a smooth recovery of the flux. As the temporal coverage of this light curve is longer than a typical orbit of the circumstellar debris producing transits, this feature is very likely the final egress of the previous B event. We also detected a third brief and shallow drop in flux of about four\,per\,cent around 140 min in both the GTC and LT light curves, but it is poorly resolved in our data. 

\begin{figure*}
	\includegraphics[width=2.11\columnwidth]{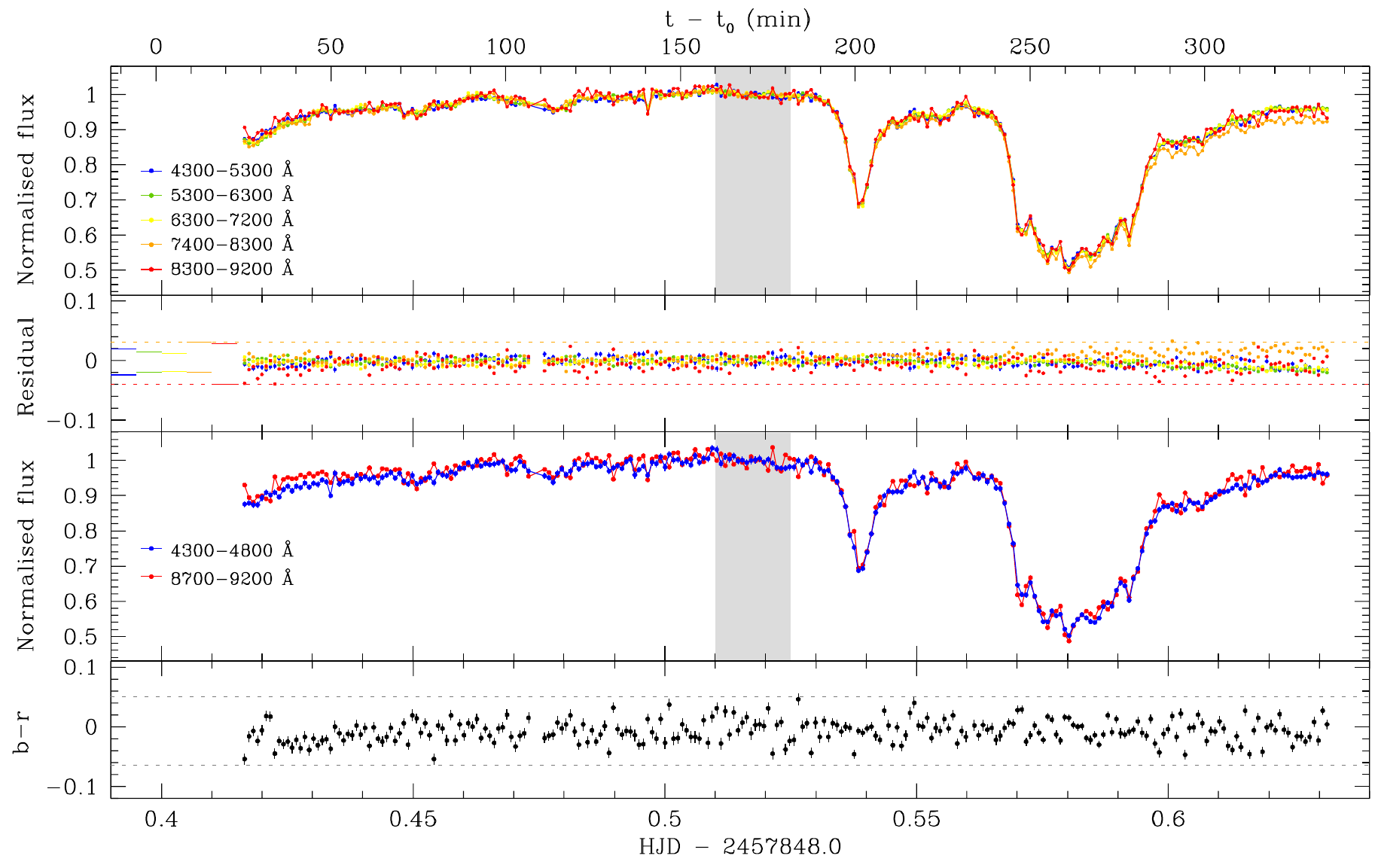}
    \caption{Top panel: GTC colour light curves constructed by integrating over five different bands. The grey shaded area encloses the points used for the normalisation. Mid-top panel: difference between the white-light light curve and the five different bands. Short horizontal lines mark the 3$\sigma$ levels and the long dashed ones the larger quantity among them. Mid-bottom panel: the bluest and reddest GTC light curves constructed by integrating over the indicated $500$--\AA\ wide bandpasses. Bottom panel: difference between the blue (b) and the red (r) light curve. Grey horizontal dashed lines mark the 3$\sigma$ level.}
    \label{fig:color_LCs}
\end{figure*}

The A transit lasts about 35\,min, reaching a maximum relative flux decrease of 30\,per\,cent, while the longer B dip lasts a total of 87\,min with a maximum depth of 50\,per\,cent. Both transits show longer egress than ingress times. A and B steeply drop to minimum light in 9 and 27\,min, and recover the 95\,per cent level in 26 and 60\,min, respectively. The final egress for both transits is much shallower than the ingress, with an apparent plateau followed by a slow recovery to maximum light after a  steep egress. In addition, the brightness level near the minimum of the B transit shows significant structure, with six individual dips that are almost equally spaced in time (see Fig.\,\ref{fig:zoom} and Sect.~\ref{sec:tr_shape} for a discussion). It is important to note that the same short-time scale variability is observed in the LT and GTC light curves. The coincidence of the same pattern of dips in the data obtained with two independent telescope/instrument combinations can be clearly seen in Fig.~\ref{fig:zoom}. This variability is wavelength-independent (see Figs.~\ref{fig:color_LCs} and~\ref{fig:ew_feii}), ruling out that it is related to red noise. Finally, the amplitude of these short-time scale structures (0.08 on average) significantly exceeds the root mean square of the out-of-transit region used for normalisation, confirming that these structures are intrinsic to \wds. 


\citet{xuetal18-1} reported time-series photometry of \wds\ that was obtained also on 2017 April 4, but covered the next orbital cycle. The A and B transits were still present in their multi-wavelength photometric observations (they labelled them as B2 and B3, respectively). While the substructure of the B dip was only marginally resolved in the light curve presented by \citet{xuetal18-1} (B3 in their figure\,2), their data do show evidence for the shallow drop in flux that we detect around $t=140$\,min. 


\subsection{GTC colour light curves}


In order to check for variations in the transit depth as a function of wavelength, we constructed light curves for a total of five different wavelength ranges, each $900-1000$\,\AA\ wide (Fig.\,\ref{fig:color_LCs}, top panel). We quantified their consistency by computing the residuals between each colour light curve and the GTC white-light light curve (Fig.\,\ref{fig:color_LCs}, mid-top panel). The five light curves behave consistently both in the out-of-transit and the in-transit regions except for a $\simeq$\,5\,per cent discrepancy in the $7400-8300$\,\AA\ light curve (orange) starting at about 275\,min. We associate this reduction in flux with 
the effect of residual telluric features in this wavelength range  and the increasing airmass ($>$1.5) towards the end of the observations. 

We reduced the wavelength coverage to extract the bluest ($4300-4800$\,\AA) and reddest ($8700-9200$\,\AA) light curves and to assess the difference between them, i.e. the colour index (Fig.\,\ref{fig:color_LCs}, mid-bottom and bottom panel, respectively). The flat colour index indicates that no significant colour effects exist over the most extreme wavelength ranges covered by the GTC data.


\subsection{Evolution of absorption lines during transits}

\begin{figure*}
    \centering
	\includegraphics[width=2.16\columnwidth]{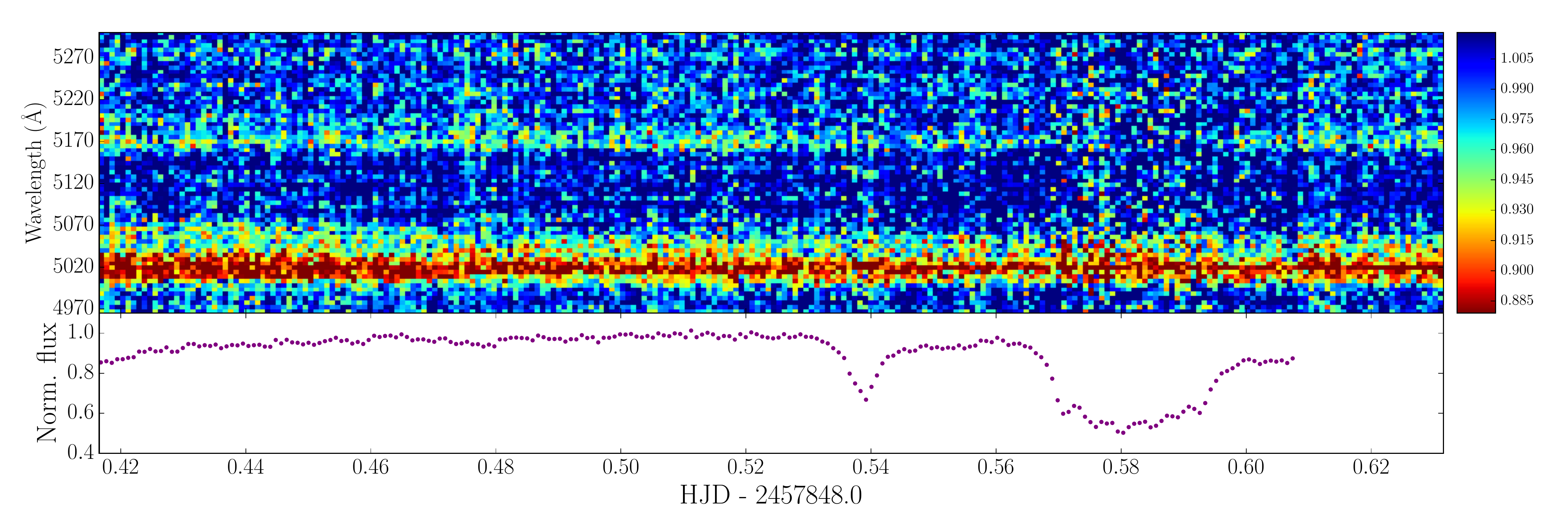}
    \caption{Top panel: Trailed spectrogram of the GTC data containing the photospheric \Line{He}{i}{5016} and the circumstellar \Line{Fe}{ii}{5169} absorption lines. The spectra have been continuum-normalised. Redder colours indicate increasing absorption. Bottom panel: Simultaneous light curve obtained with the LT. \Line{Fe}{ii}{5169} clearly weakens during the B transit, whereas the \Line{He}{i}{5016} line remains largely unaffected. Note that the signal-to-noise ratio decreases during the transits.}
    \label{fig:trail}
\end{figure*}

The spectrum of \wds\ contains broad \Ion{He}{i} absorption lines, as well as multiple narrower metal lines that are both of photospheric and circumstellar origin \citep{xu16-1}. The circumstellar lines exhibit short-term variability potentially associated with transit events \citep{redfieldetal17-1}. We constructed a trailed spectrogram of the strongest circumstellar absorption line contained in the wavelength range of the GTC data, \Line{Fe}{ii}{5169} (Fig.\,\ref{fig:trail}). It clearly shows, contrary to the photospheric \Line{He}{i}{5016} line, a reduction in strength during the B transit. 

\begin{figure*}
	\includegraphics[width=2.12\columnwidth]{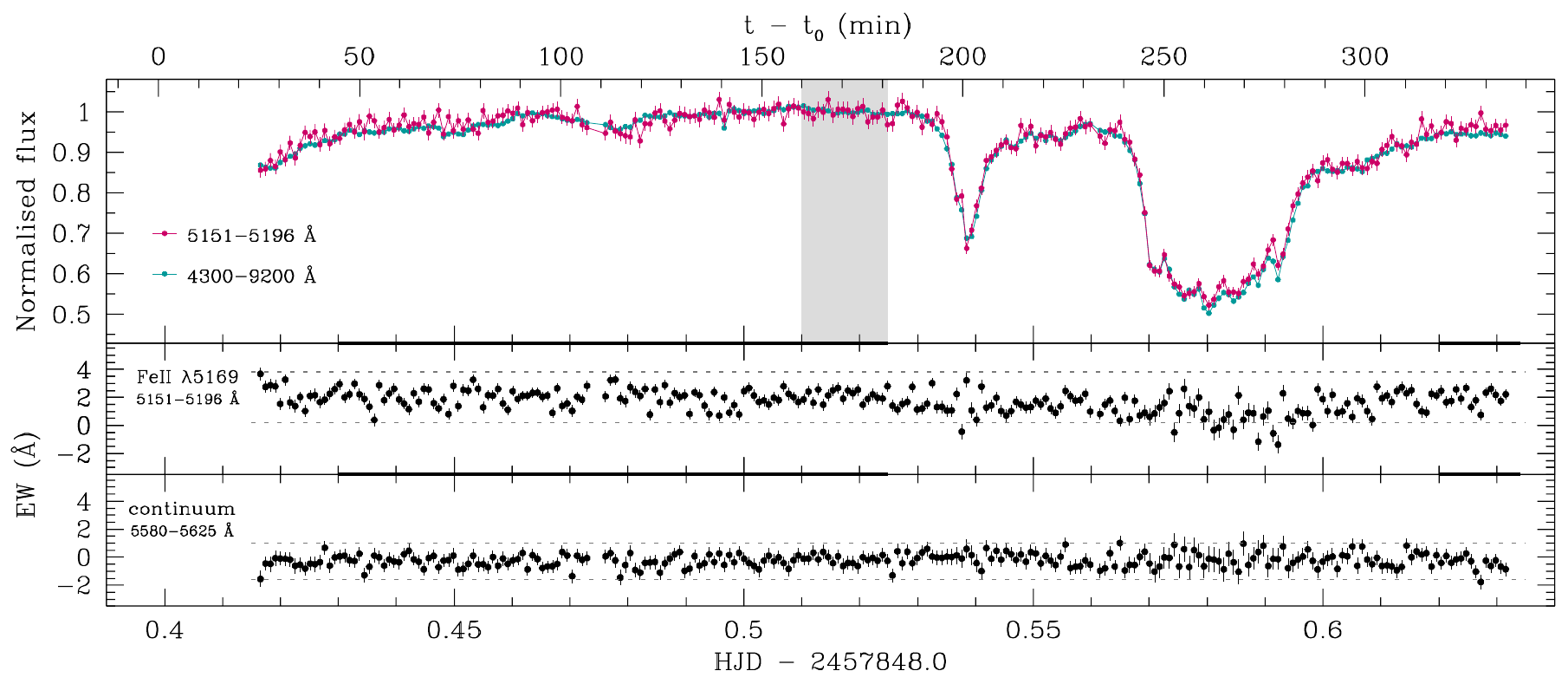}
    \caption{Top panel: light curves produced by integration over the wavelength ranges shown in the legend. The grey shaded area indicates the range used for the normalisation. Mid and bottom panels: equivalent width of the circumstellar \Line{Fe}{ii}{5169} absorption line and of the surrounding continuum. The grey dashed lines mark the $3 \sigma$ level of each out-of-transit distribution, which are defined by the thick black lines on both top-axes.}
    \label{fig:ew_feii}
\end{figure*}

To quantify the changes in the strength of the circumstellar \Line{Fe}{ii}{5169} absorption line, we computed its equivalent width for all individual GTC spectra over the wavelength range $5151-5196$\,\AA, as well as for a neighbouring continuum region between $5580-5625$\,\AA. During the B transit the equivalent width of the \Line{Fe}{ii}{5169} line falls significantly below its out-of-transit mean value, whereas the one of the neighbouring continuum remains unaffected (Fig.\,\ref{fig:ew_feii}, mid and bottom panel, respectively). As a final, visual check we compare in Fig.\,\ref{fig:ew_feii} (top panel) the light curve of the \Line{Fe}{ii}{5169} line, also computed over $5151-5196$\,\AA, with the white light light curve, which shows a deeper transit during minimum light of the B transit.

\begin{figure}
    \centering
    \includegraphics[width=0.8\columnwidth]{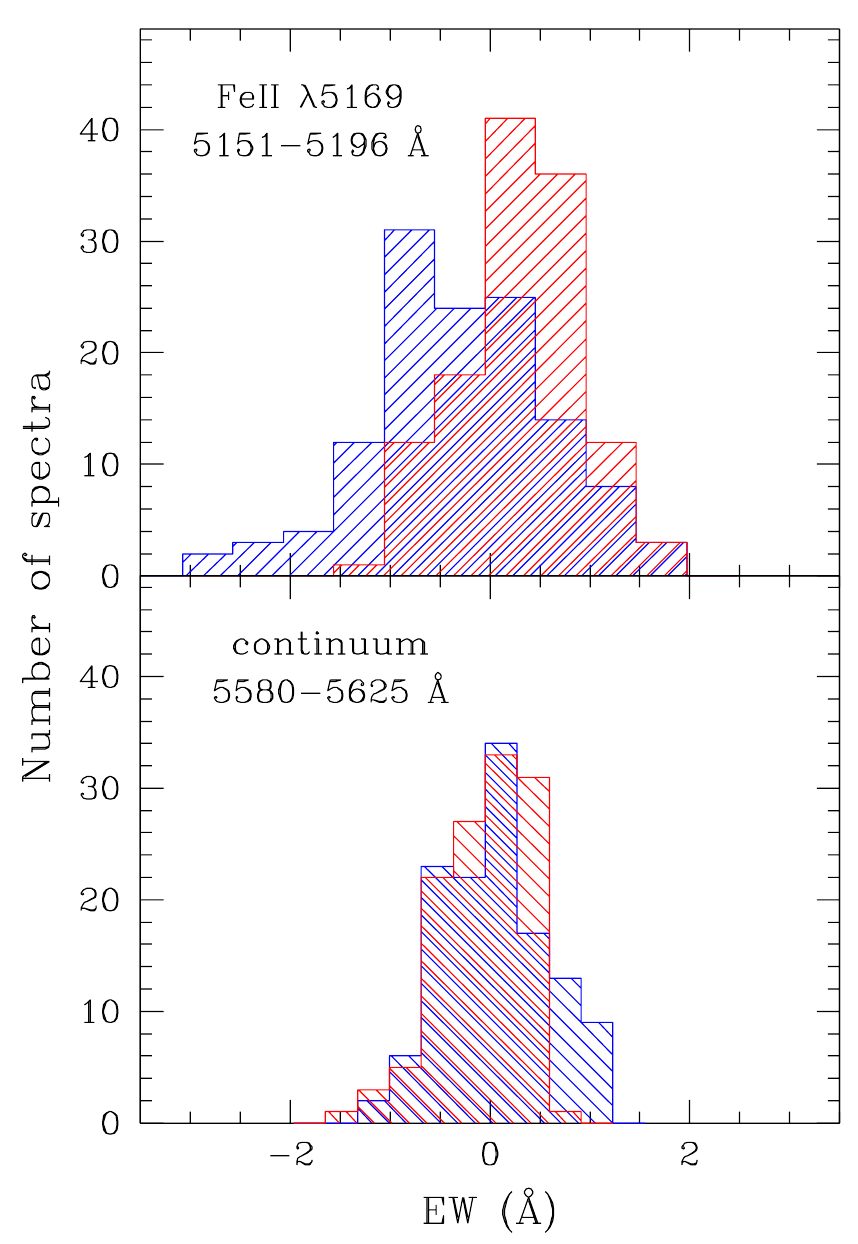}
    \caption{Equivalent width of the circumstellar \Line{Fe}{ii}{5169} absorption line (top) and its surrounding continuum (bottom). The blue histograms correspond to the in-transit spectra and the red ones to the out-of-transit spectra. The total mean of both sets of distributions has been subtracted (1.71\,\AA\ for the \Line{Fe}{ii}{5169} line and $-0.22$\,\AA\ for the continuum).  }
    \label{fig:histo}
\end{figure}

We computed two histograms of the equivalent width of the circumstellar \Line{Fe}{ii}{5169} line and two for the neighbouring continuum region (Fig.\,\ref{fig:histo}). For each region, we produced one histogram for the in-transit spectra (points below 44.7\,min or between 181.5 and 318\,min, blue histograms), and another one for the out-of-transit points (which contain all other points, red histograms). The mean of the whole distribution (the in-transit plus the out-of-transit points) has been subtracted from each pair of histograms.  

The two \Line{Fe}{ii}{5169} histograms (Fig. \ref{fig:histo}, top panel) are clearly distinct. The out-of-transit points (red histogram) show a Gaussian distribution with the peak centred on positive equivalent widths, i.e. absorptions. However, the in-transit spectra (blue histogram) show a significant trend towards negative equivalent widths, with an enhanced tail that makes the distribution asymmetric. This dichotomy is not present in the histograms for the continuum (Fig. \ref{fig:histo}, bottom panel), where the in-transit and out-of-transit points behave consistently, with similar, symmetrical distributions. 



\section{Discussion}

\subsection{Duration and shape of the transits}
\label{sec:tr_shape}


The transit signatures observed in \wds\ as part of the intense follow-up observations obtained since their discovery \citep{vanderburg15} share multiple characteristics with the candidate disintegrating planets KIC\,12557548\,b \citep{rappaport12}, KOI\,2700-b \citep{rappaport14} and EPIC\,201637175 \citep{sanchis-15}. All have short orbital periods ($\sim$few hours) and exhibit variable transit depths and asymmetric transit shapes.

The two transits reported here, A (30\,per cent depth, see Fig.\,\ref{fig:lt_gtc_LCs}) and B (50\,per cent) were already present at least two weeks before our observations (2017 March~22, see \citealt{rappaportetal18-1}), and during that time the transits evolved both in depth and shape. The transit durations we measured (35 and 87\,min for A and B, respectively) vastly exceed the expected time for a solid body transiting the white dwarf ($\simeq1$\,min). Both occultations display a sharp ingress and a longer egress, with a final plateau and a gradual recovery to maximum light. The evolving, asymmetric shapes of the transits and their long durations are consistent with the hypothesis of \citet{vanderburg15} of being the signatures of dust and gas clouds emanating from solid fragments on a $\simeq4.5$\,h orbital period, with cometary-like tails trailing the occulting material.

\begin{figure}
	\includegraphics[width=\columnwidth]{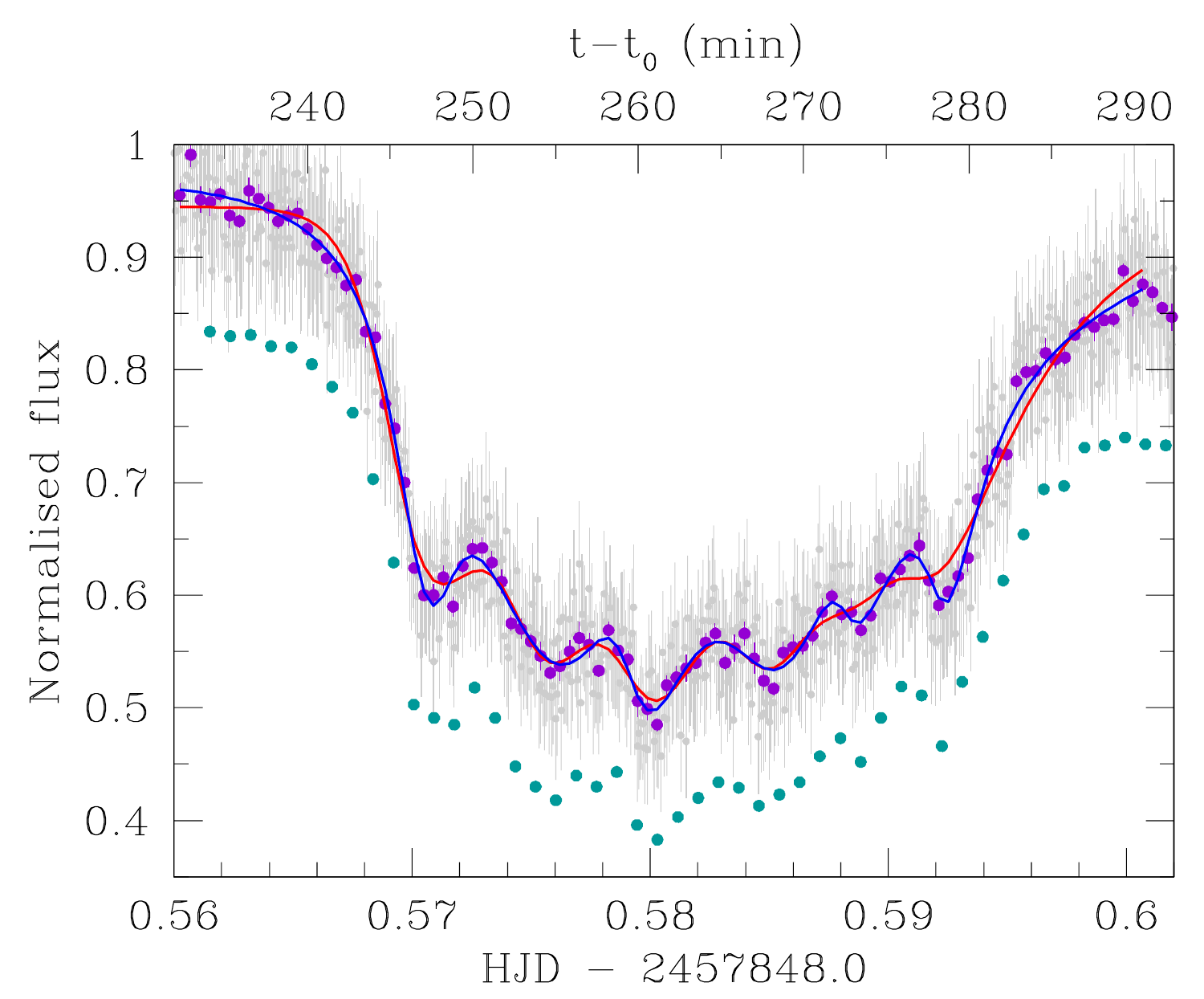}
    \caption{The LT light curve at the original 5-s cadence (gray) and binned by a factor seven (purple points) and zoomed in the B transit. The solid lines are the best fits with the function given by Eq.~\ref{eq_fit}, where the ingress and egress time scales have been fixed to the values given by a fit to the A dip (red), or considered as free parameters (blue). The GTC white-light light curve (green points) has also been plotted with an offset $-0.1$ in order to emphasise the match between the short-term variability in the two independent data sets. The top $x$-axis ($t-t_{0}$) is the time in minutes from the first data point of the unbinned LT light curve.}
    \label{fig:zoom}
\end{figure}

In addition, we detected a complex structure at minimum brightness of the B transit. We propose two distinct explanations for the large duration of the B transit and its substructure: the superposition of dust clouds emanating from multiple solid fragments that are nearly equally spaced in azimuth, or density structures within the extended dust tail associated with a single fragment.

In the first scenario, the individual dips detected in the light curve could be associated with fragments spaced $\simeq10^\circ$  apart in azimuth. \citet{salo+yoder88-1} investigated the evolution of up to nine co-orbital bodies, and found stable configurations where all bodies were clumped on one side of the orbit. Dynamical  stability of six co-orbital fragments was explored in the context of \wds~ \citep{gurrietal17-1}, finding that they remain stable for up to two years as long as their masses are $\la10^{23}$\,g and their eccentricities less than 0.1. A more general quantification of the stability of co-orbital configurations of objects orbiting white dwarfs has been studied \citep{veras16-01}, obtaining results which suggest that stability may be achieved with different numbers of co-orbital, Ceres-like objects. Although none of these studies included fragments as closely spaced as $\simeq10^{\circ}$, it appears plausible that sufficiently low fragment masses can remain in stable orbits for such a configuration. 

In order to test the viability of this scenario, we follow \citet{rappaport14} in using the asymmetric hyperbolic secant (AHS) function to fit the ingress and egress times of the transits individually. However, instead of adopting it as an additive term, we use the AHS function as a multiplicative transmission function, which easily allows to combine the effect of multiple overlapping dips as: 

\begin{equation}
   \mathrm{Flux}(t)=f_{0}\ \prod_{n} \bigg( \frac{a_{n}}{e^{-(t-t_{n})/t_{i,n}}+e^{(t-t_{n})/t_{e,n}}}+1 \bigg)^{-1}
   \label{eq_fit}
\end{equation}
\\

\noindent where $f_{0}$ is the out-of-transit flux level, $n$ represents the number of AHS and $a_{n}$, $t_{n}$, $t_{i,n}$ and $t_{e,n}$ are the depth factor scale, the time at minimum flux and the ingress and egress time scales of the $n-$dip, respectively. Therefore, the fitting method used here has $4n+1$ free parameters and we used a Levenberg-Marquardt algorithm \citep{levenberg44,marquardt63} to derive the least-squares fit.

We first fitted the A transit with Eq.~\ref{eq_fit} and $n=1$ (five free parameters) and obtained $t_{i,1}(\mathrm{A})=1.5$ and $t_{e,1}(\mathrm{A})=5.6$\,min. We then fitted the B transit with the superposition of $n=6$ A-transit profiles, i.e. keeping $t_{i,n}$ and $t_{e,n}$ fixed to the values found for the A dip, and leaving $f_{0}$, $a_{n}$ and $t_{n}$ free (a total of 13 free parameters). This approach reproduced reasonably well the first four minima in the B dip, but failed to match the last two minima (see the red curve in Fig.\,\ref{fig:zoom}). As a second experiment, we performed a fit to the B transit again with $n=6$ transit profiles, but leaving also $t_{i,n}$ and $t_{e,n}$ as free parameters (hence a total of 25 free parameters, blue line in Fig.\,\ref{fig:zoom}).  This model provides a much better fit to the substructure observed in dip B. The parameters of the fits to the A and B transits are reported in Table~\ref{tab:fits}.

Both fits show that the broad B transit can be reproduced by the superposition of six individual transits caused by multiple fragments closely spaced in azimuth. There is no reason for the six fragments to produce identical ingress and egress time scales, yet the overall agreement of the 13-parameter fit is remarkably good. In contrast, while the 25-parameter fit results in a much closer agreement with the observed shape of the B transit, the larger uncertainties in the fit parameters suggest that the increased number of free parameters is under-constrained. We note in passing that the six times of transit minimum are roughly equally spaced in time, with an average separation of $6.2\pm1.1$\,min. The nearly equal spacing of these bodies is reminiscent of the string of fragments of comet Shoemaker-Levy~9 \citep{weaveretal94-1}.

\begin{table*}
  \centering
\begin{tabular}{cccccccc}
\hline
\hline
Dip & $\#$ free parameters & $n$ & $f_{0}$ & $a_n$  & $t_{n}$ (min)  & $t_{i,n}$ (min) & $t_{e,n}$ (min)  \\
\hline
A & $5$ & $1$ & $0.981 \pm 0.005$  & $0.63 \pm 0.05$ & $198.5 \pm 0.4$ & $1.5 \pm 0.2$ & $5.6 \pm 0.6$ \\
\hline
B & $13$ & $1$ & $0.945 \pm 0.004$   & $0.88 \pm 0.03$ & $246.02 \pm 0.12$ & $1.5$ & $5.6$\\
  &      & $2$ &                     & $0.74 \pm 0.04$ & $253.4 \pm 0.2$ & $1.5$ & $5.6$\\
  &      & $3$ &                     & $0.78 \pm 0.05$ & $259.9 \pm 0.2$ & $1.5$ & $5.6$\\
  &      & $4$ &                     & $0.62 \pm 0.05$ & $266.9 \pm 0.3$ & $1.5$ & $5.6$\\
  &      & $5$ &                     & $0.31 \pm 0.05$ & $272.3 \pm 0.8$ & $1.5$ & $5.6$\\
  &      & $6$ &                     & $0.36 \pm 0.05$ & $277.1 \pm 0.4$ & $1.5$ & $5.6$\\
\hline
B & $25$ & $1$ & $0.968 \pm 0.010$ & $0.64 \pm 0.05$ & $247.0 \pm 0.3$ & $1.4 \pm 0.3$ & $1.3 \pm 0.3$\\
  &      & $2$ &                   & $1.58 \pm 0.06$ & $255.4 \pm 1.2$ & $4.5 \pm 0.7$ & $3.7 \pm 2.3$\\
  &      & $3$ &                   & $0.53 \pm 0.37$ & $259.4 \pm 0.4$ & $0.7 \pm 0.4$ & $11 \pm 18$\\
  &      & $4$ &                   & $0.72 \pm 0.50$ & $267.5 \pm 1.8$ & $2.0 \pm 1.0$ & $5 \pm 5$\\
  &      & $5$ &                   & $0.21 \pm 0.16$ & $272.3 \pm 0.2$ & $0.5 \pm 0.4$ & $15 \pm 27$\\
  &      & $6$ &                   & $0.38 \pm 0.06$ & $278.0 \pm 0.4$ & $0.8 \pm 0.3$ & $1.7 \pm 0.6$\\
\hline
\end{tabular}
\caption{Parameters of the best fits to transits A and B using Eq.~\ref{eq_fit}. The 13-free-parameter fit to the B dip was computed fixing the $t_{i,n}$ and $t_{e,n}$ to the values obtained for the A dip.}
\label{tab:fits}
\end{table*}

The alternative scenario of a single dust cloud was  proposed before by \citet{alonso16-1} to explain the structure of the transits observed on 2016 January 18 by changes in the optical depth of the debris. The B transit extends $\simeq120^\circ$ in azimuth, and it is not clear what mechanism would cause the relatively sharp ingress and egress of such a large debris cloud associated with a single solid fragment.


\subsection{Grey transits\label{sec:grey}}

We extracted light curves over five different bands from the GTC data, finding no significant differences in the depth of the transits as a function of wavelength (see Fig. \ref{fig:color_LCs}). Following the detection of the `bluing' \citep{hallakounetal17-1}, we carefully compared the bluest and reddest light curves extracted from the GTC data, again not finding any difference beyond the $2\sigma$ level at any interval. However, we were limited in the spectral range of the GTC observations to the equivalent of the $g$-band, whereas they obtained $u$-, $g$-, $r$-, and $i$-band photometry. 

In the canonical optically-thin scenario, this result is in good agreement with the findings of \citet{alonso16-1}. They argued that the transits in \wds\ are grey across the optical range, and deduced a lower size of $0.5\,\mathrm{\mu m}$ for the dust particles by computing the \AA ngstr\"om exponents for common minerals. Similar conclusions were reached  based on observations obtained from 2015 to 2017 \citep{zhouetal16-1,crolletal17-1,xuetal18-1}. Despite the rapid evolution in the number, depths, and shapes of the transits suggesting the dynamical evolution of the system, there is as yet no sign of the production of small particles.

However, in the optically-thin scenario discussed so far in the published studies, the dust cloud responsible for the transits must cover a great part of the height of the white dwarf as seen from Earth. By optically thin we understand an optical depth $\tau\ll1$ (or at least $\tau<1$), since $\tau=1$ corresponds to an attenuation along that line of sight of $(1-e^{-1})=63$\,per\,cent. Consequently, an optically-thin medium would have to extend roughly the whole vertical extent of the stellar disc to get a transit depth of 50\,per\,cent. The dust grains would therefore have a significant vertical velocity dispersion and experience collisions that would result in the generation of yet smaller dust grains. 


Taking a white dwarf mass of $0.636\mathrm{\,M}_\odot$ (Table~\ref{tab:params}), the orbital velocity corresponding to a 4.5\,h orbit is 320\,km\,s$^{-1}$. With a white dwarf radius of $0.01205\mathrm{\,R}_\odot$ (Table~\ref{tab:params}), an optically thin dust cloud covering the disc of the star must comprise grains with orbital inclinations up to $i\approx0.58^\circ$ with respect to the line of sight, and corresponding vertical velocities up to $v_z\approx3.25\mathrm{\,km\,s}^{-1}$. Using a cloud radius of $9.5\mathrm{\,R}_\oplus$ as postulated by \cite{crolletal17-1}, these numbers increase to $i\approx4.30^\circ$ and $v_z\approx23.50\mathrm{\,km\,s}^{-1}$. By comparison, the collision velocity needed for catastrophic fragmentation of $1\,\mu\mathrm{m}$ grains is only $0.12\mathrm{\,km\,s}^{-1}$ \citep{stewart+leinhardt09-1}. Thus, ongoing generation of small grains is unavoidable.


As well as the collision velocities, the collision rate is also a factor in the production of small grains. Owing to the relatively large optical depth of this ``optically-thin" cloud (a 50 per\,cent transit depth corresponds to an optical depth along the radial line-of-sight of $\tau_\mathrm{rad}=0.7$), grain--grain collisions will occur frequently. If the dust cloud has a circular cross-section, then the vertical optical depth is also $\tau_\mathrm{vert}=0.7$. Grains in the cloud traverse it vertically twice per orbit (once ascending, once descending), meaning that only 25 per\,cent of grains will not have suffered a collision in a single 4.5\,h orbit. However, the vertical optical depth can be reduced if the dust cloud is more extended radially than vertically. If the radial extent is $k$ times the vertical extent, the half-life of the dust grains (in units of the orbital period) is $N_{1/2}=k\ln2/2\tau_\mathrm{rad}$. Now, the observed lifetime of transit features ranges from $\sim$days to $>$months \citep[e.g.][]{gansicke16-1}. Taking a minimum of ten orbits for the lifetime of the transit features implies $k>20$ to avoid significant collisional evolution of grains over the lifetime of the cloud. Such a large value of $k$ corresponds to a radial extent of the dust features of $50\,\mathrm{\,R}_\oplus$, or $0.4$ times the orbital radius, which would rapidly destroy such physical structures through the action of Keplerian shear. To maintain the observed coherence of the transit features, $k$ must be small, resulting in a rapid production of small grains, fast enough that they should be detectable.

While the observed lack of small grains may be explained by their short lifetimes to sublimation \citep{xuetal18-1}, the origin of the vertical motions remains to be explained. A parent planetesimal can only gravitationally excite vertical velocities comparable to its escape velocity, i.e. $0.65\mathrm{\,km\,s}^{-1}$ for a 500\,km radius and a density of 3\,g\,cm$^{-3}$, insufficient to excite the cloud to cover the face of the white dwarf. An alternative source of velocity excitation would be turbulent motions in the gas, if the gas disc is sufficiently dense and the grains well coupled. With grain temperatures ranging from 1100 to 1900\,K \citep{xuetal18-1}, thermal velocities of oxygen atoms (a light species) in the gas would be 1.36 to 1.76\,km\,s$^{-1}$. Hence, not only would the grains have to be well-coupled to a turbulent gas disc, but the turbulence would have to be highly supersonic.

The above considerations motivate us to reconsider an optically-thick, geometrically-thin cloud. We explore this scenario in Appendix~\ref{ap:optical}, finding that it is not only possible but also viable.  The inclinations for the optically-thick discs are $\sim 0.08^\circ$, consistent with stirring by a 500\,km planetesimal. In this case, a wavelength dependence of the observed transit depth is prevented by the large gradient of optical depth moving away from the mid-plane (see Fig.~\ref{fig:optically_thick}, top). Inferred dust masses lie between the estimated mass of the \Ion{Fe}{ii} component of the gas disc \citep{redfieldetal17-1}, and the total mass of accreted material and mass of the parent body \citep[][ Fig.~\ref{fig:optically_thick}, bottom]{redfieldetal17-1}. The high optical depth in the mid-plane would in fact shield the small grains from sublimation. A potential challenge to this model, however, is that it cannot explain the IR excess, which then requires an additional, inclined, dust component \citep{xuetal18-1}.

\subsection{Variability of circumstellar absorptions during transit}

The trailed spectrogram of one broad photospheric \Ion{He}{i} absorption line ($\lambda 5016$) and the strongest circumstellar line in our spectral coverage, \Line{Fe}{ii}{5169}, clearly shows the latter almost disappearing during the B transit, while \Line{He}{i}{5016} remains steady. We measured the equivalent width evolution of the \Ion{Fe}{ii} line (see Fig. \ref{fig:ew_feii}) and found a decrease and subsequent increase, which match the ingress and egress times of the photometric transit.

A decrease in the strength of some circumstellar features associated with major photometric transits has previously been suggested by \citet{redfieldetal17-1}, though their data were lacking simultaneous photometry. Our wide-slit GTC spectrophotometry, accompanied by the simultaneous LT photometry, unambiguously confirms the correlation between continuum extinction and the equivalent width decrease of the circumstellar absorption lines.


In order to see a decrease in the strength of the circumstellar line, a fraction of the light that passes through the gas and produces the absorption has to be blocked, while the light that does not interact with this gas is not, or it is less affected. The qualitative implications derived from this are: (1) the gas is not covering the full projected stellar disc of the white dwarf, (2) during the B transit, the dust blocks a larger fraction of the light that penetrates the circumstellar gas than that unaffected by the gas. We develop in Appendix~\ref{sec:mark} a simple model that quantitatively accounts for the fraction of the white dwarf covered by dust and/or gas and its implications on the depth of the circumstellar lines. The main conclusions of this model are that the fraction of the gas covered by dust is around $1.5$--$2$ times greater than the fraction of the star that is covered only by dust  (in order to see the reported decrease in the line), and that the weakening of the circumstellar absorption lines due to dust would happen independently of the relative location of gas and dust (the gas being closer to the white dwarf than the dust, or vice versa), as was previously suggested \citep{hallakounetal17-1}.

\section{Conclusions}

We have obtained simultaneous fast spectrophotometry and photometry of \WD\ spanning more than one orbital cycle of the major debris fragments, and detected two deep transits which both showed a remarkable delay in the egress compared to their ingress times, consistent with the scenario of a comet-like tail emanating from the fragments.

The deepest transit shows substantial substructure near minimum light caused either by a closely spaced chain of rocky fragments, akin to the debris stream of comet Shoemaker-Levy 9, or by a debris cloud extending $120^\circ$ in azimuth with an inhomogeneous density structure. 

Extracting five $900-1000$\,\AA\ wide light curves from the GTC spectrophotometry reveals no significant colour effects in the depth of the transits, confirming the grey nature of the dust that has been established by multiple studies based on data taken from 2015 to 2017. We explored the viability of an optically-thick scenario in order to account for the non-detection of small particles, and concluded that it is possible and simpler than the optically-thin one. 


We have detected a significant decrease in the strength of the circumstellar \Line{Fe}{ii}{5169} absorption line during the deepest transit, where the association is confirmed by our simultaneous broadband photometry. This result strongly suggests that gas and dust are spatially correlated. During the deepest transit, dust blocks a larger fraction of light that would otherwise interact with the circumstellar gas than light unaffected by gas. The decrease in the strength of the circumstellar line provides no information on the relative location of gas and dust along the line-of-sight towards the white dwarf. Time-resolved spectroscopy with a higher signal-to-noise ratio has the potential to provide further insight into the geometry of the gas and dust within \WD. 

Finally, we used the parallax obtained from \textit{Gaia} Data Release 2 and photometric data of three independent surveys to derive the atmospheric parameters of \WD. 
Adopting the most likely reddening, $E(B-V)=0.01$, results in $T_\mathrm{eff}=15020 \pm 520$\,K and $\log g=8.07\pm0.07$ for a DBAZ composition of the atmosphere, which correspond to $M_\mathrm{WD}=0.63\pm0.05\,\Msun$ and a cooling age of $224\,\pm30$\,Myr.

\section*{Acknowledgements}
We wish to thank Roi Alonso and Rik van Lieshout for helpful discussions. We would also like to thank the referee for a careful reading of the manuscript and helpful feedback. PI acknowledges financial support from the Spanish Ministry of Economy and Competitiveness (MINECO) under the 2015 Severo Ochoa Programme MINECO SEV--2015--0548. The research leading to these results has received funding from the European Research Council under the European Union's Seventh Framework Programme (FP/2007--2013) / ERC Grant Agreement n.\,320964 (WDTracer). OT was supported in part by a Leverhulme Trust Research Project Grant. AJM is supported by the project grant 2014.0017 `IMPACT' from the Knut \& Alice Wallenberg Foundation, and by the starting grant 2017--04945 `A unified picture of white dwarf planetary systems' from the Swedish Research Council. This research has been supported by the Jet Propulsion Laboratory through the California Institute of Technology postdoctoral fellowship program, under a contract with the National Aeronautics and Space Administration. RR acknowledges funding by the German Science foundation (DFG) through grants HE1356/71-1 and IR190/1-1. DV gratefully acknowledges the support of the Science and Technology Facilities Council via an Ernest Rutherford Fellowship (grant ST/P003850/1). This project has received funding from the European Research Council (ERC) under the European Union's Horizon 2020 research and innovation programme (grant agreements No. 677706--WD3D). Based on observations made with the Gran Telescopio Canarias (GTC), installed in the Spanish Observatorio del Roque de los Muchachos of the Instituto de Astrof\'isica de Canarias (IAC), in the island of La Palma, and on observations made in the Observatorios de Canarias of the IAC with the Liverpool Telescope (LT) operated on the island of La Palma by the Liverpool John Moores University in the Observatorio del Roque de los Muchachos. Data for this paper have been obtained under the International Time Programme of the CCI (International Scientific Committee of the Observatorios de Canarias of the IAC).




\bibliographystyle{mnras}
\bibliography{biblio} 



\appendix

\section{An optically-thick cloud}
\label{ap:optical}

The scenario of an optically-thin dust cloud causing the observed transits presents two main problems: the unknown origin of particle vertical motions and the lack of small particles (the non-detection). Therefore, we explored an alternative scenario: an optically-thick and geometrically-thin cloud. 

For simplicity, we model this as a rectangular profile crossing the white dwarf, with height $2h_\mathrm{r}$ and $2h_\mathrm{b}$ in red ($\lambda_\mathrm{r}=8950$\,\AA) and blue ($\lambda_\mathrm{b}=4550$\,\AA) wavebands, respectively. We consider the cloud to be totally opaque below $|z|=h$ and transparent above $|z|=h$, identifying $|z|=h$ with the surface where the optical depth $\tau$ is $1$ (as we shall see, the vertical gradient in optical depth is large, so this approximation is valid). 

From the lack of observed wavelength dependence on the transit depth $D$, we have a $3\sigma$ upper limit of $|D_\mathrm{b}-D_\mathrm{r}|\lesssim0.06$ on the difference between red and blue transit depths (Fig~\ref{fig:color_LCs}). Taking into account that
\begin{equation}
\Delta D = \frac{2 (h_b-h_r) \sqrt{R_{WD}^2 - h_r^2}}{\pi R_{WD}^2},
\end{equation}
this corresponds to a difference of $|h_\mathrm{b}-h_\mathrm{r}|\lesssim0.13\mathrm{\,R}_\oplus$ on the heights on the $\tau=1$ surface. We now demonstrate that an optically-thick cloud composed of small grains (with very different opacities which would be detectable in the optically-thin case) can indeed satisfy this constraint and remain undetectable, so long as it is sufficiently concentrated in the mid-plane and the vertical density gradient is sufficiently high.

If the grains have a Rayleigh distribution of inclinations with parameter (mode) $\sigma_I$ (appropriate for random excitations), then the vertical number density $n$ and optical depth are Gaussian:
\begin{equation}
	\tau \propto n \propto \frac{\exp\left(-z^2/2a^2\sigma_I^2\right)}{\sqrt{2\pi}a\sigma_I},
\end{equation}
where $a$ is the orbital semi-major axis. In particular, the mid-plane optical depth is given by
\begin{equation}
	\tau_{0,\mathrm{r}} = \exp\left(h_\mathrm{r}^2/2a^2\sigma_I^2\right).
\end{equation}
The ratio of optical depths in the two wavebands can be parameterised by the {\AA}ngstr\"om exponent $\alpha$:
\begin{equation}
	\tau_\mathrm{b}/\tau_\mathrm{r} \propto \left(\lambda_\mathrm{r}/\lambda_\mathrm{b}\right)^\alpha,
\end{equation}
where typically $\alpha$ lies between 1 and 4. Then $h_\mathrm{b}$ satisfies
\begin{equation}
	1 = \left(\frac{\lambda_\mathrm{r}}{\lambda_\mathrm{b}}\right)^\alpha \tau_{0,\mathrm{r}}\exp{\left(-h_\mathrm{b}^2/2a^2\sigma_I^2\right)}
\end{equation}
or, if $|h_\mathrm{b}-h_\mathrm{r}|\ll h_\mathrm{r}$,
\begin{equation}
	h_\mathrm{b} = h_\mathrm{r}\left[1 + \alpha\left(\frac{a\sigma_I}{h_\mathrm{r}}\right)^2\ln\left(\frac{\lambda_\mathrm{r}}{\lambda_\mathrm{b}}\right)\right].
\end{equation}
The corresponding change in observed transit depth between the two wavebands is therefore
\begin{equation}
	D_\mathrm{b}-D\mathrm{r} = \frac{2\alpha}{\pi}\ln\frac{\lambda_\mathrm{r}}{\lambda_\mathrm{b}}\sigma_I^2\frac{h_\mathrm{r}}{R_\mathrm{WD}}\left(\frac{h_\mathrm{r}}{a}\right)^{-2}\sqrt{1-\left(\frac{h_\mathrm{r}}{R_\mathrm{WD}}\right)^2}.
\end{equation}
Given an observed limit on this change $\Delta D$, we therefore have an \emph{upper} limit on the inclination parameter
\begin{equation}
	\sigma_I^2 \le \frac{\pi\Delta D}{2\alpha\ln\left(\lambda_\mathrm{r}/\lambda_\mathrm{b}\right)}\frac{R_\mathrm{WD}}{h_\mathrm{r}}\left(\frac{h_\mathrm{r}}{a}\right)^2\left[1-\left(\frac{h_\mathrm{r}}{R_\mathrm{WD}}\right)^2\right]^{-1/2},
\end{equation}
and a corresponding \emph{lower} limit on the mid-plane optical depth
\begin{equation}
	\tau_\mathrm{0,r} \ge \exp\left[\frac{\alpha\ln\left(\lambda_\mathrm{r}/\lambda_\mathrm{b}\right)}{\pi\Delta D} \frac{h_\mathrm{r}}{R_\mathrm{WD}} \sqrt{1-\left(\frac{h_\mathrm{r}}{R_\mathrm{WD}}\right)}\right].
\end{equation}
\\

\begin{figure}
	\includegraphics[width=0.44\textwidth]{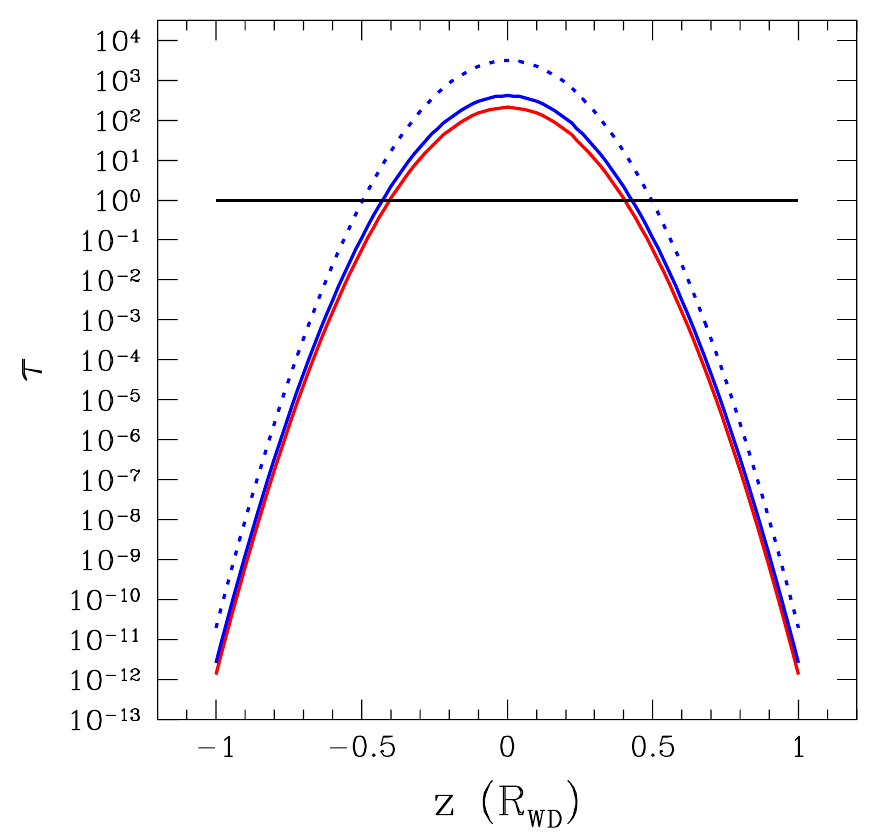}
    \includegraphics[width=0.49\textwidth]{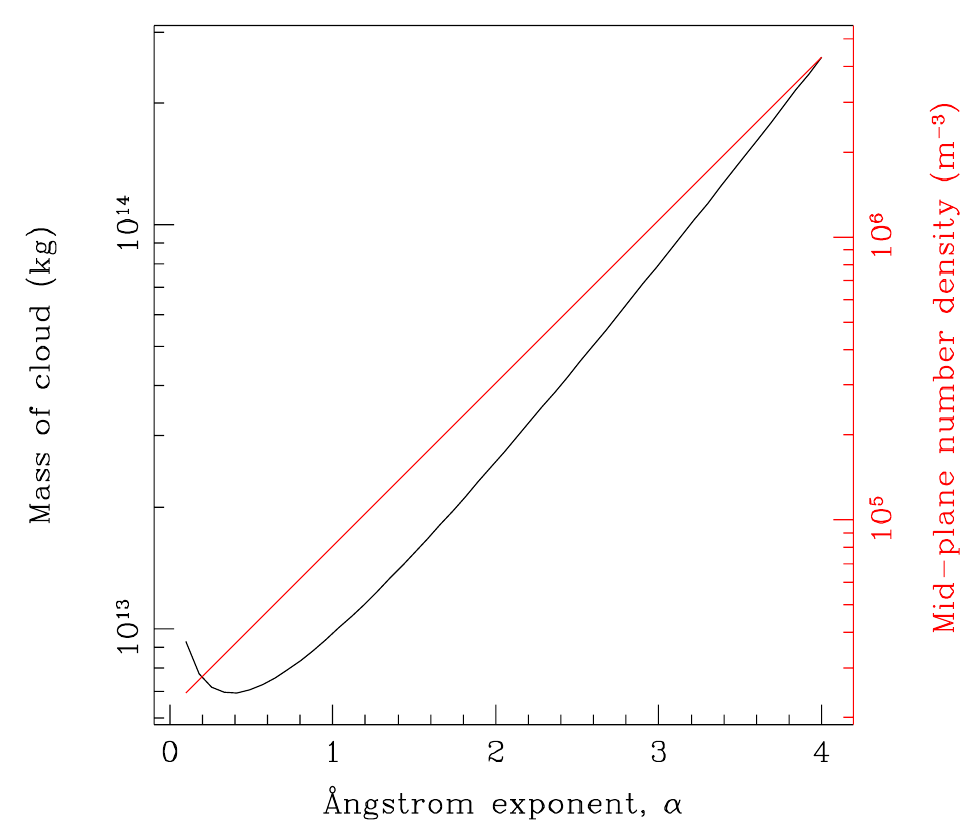}
    \caption{Top: Optical depth $\tau$ of the geometrically-thin dust cloud as a function of height $z$, with an inclination parameter $\sigma_I=0.08^\circ$. The red, blue and dashed blue lines correspond to the optical depths of the red and blue  ($\alpha=1$ and $\alpha=4$) wavebands, respectively. Moving away from the mid-plane, the optical depth transitions rapidly from $\tau\gg1$ to $\tau\ll1$. Only the narrow region between the red and the blue curves would contribute to an observed difference in transit depth between the red and the blue wavebands. Bottom: total mass in small grains and mid-plane number density for a range of {\AA}ngstr\"om exponents $\alpha$.}
    \label{fig:optically_thick}
\end{figure}

As an example, with $\alpha=4$ and the observed limit $\Delta D < 0.06$, we have $\sigma_I<0.072^\circ$ and $\tau_{\mathrm{0,r}}>200$. The corresponding vertical profile in optical depth is shown in Fig.~\ref{fig:optically_thick}. 

As a sanity check, we now compute the corresponding total mass and mid-plane number density of grains, assuming a grain size of $0.3\,\mu\mathrm{m}$. We find a number density of $4.3\mathrm{\,cm}^{-3}$ and a total mass $2.6\times10^{17}$\,g. This compares with an estimate of the total mass accreted onto the white dwarf of 2--$3\times10^{23}$\,g and a mass of gaseous \ion{Fe}{ii} of $>10^{11}$\,g \citep{redfieldetal17-1}, and with a mass of $1.6\times10^{24}$\,g for a 500\,km parent body. Indeed, this parent body would have a surface escape velocity $0.65\mathrm{\,km\,s}^{-1}$, which now compares favourably with the velocity dispersion of the geometrically thin cloud ($0.40\mathrm{\,km\,s}^{-1}$).

The {\AA}ngstr\"om exponent $\alpha=4$ is the most extreme value in the Mie scattering regime, with the largest difference in opacity between red and blue wavebands, and therefore gives the strongest constraint on vertical flattening and the mid-plane optical depth. The inferred cloud masses and densities are lower for lower $\alpha$, with the cloud mass only $10^{16}$\,g if $\alpha=1$. The dependence of mass and number density on $\alpha$ is shown in the lower panel of Fig.~\ref{fig:optically_thick}.

\section{Geometrical model for gas absorption during transit}
\label{sec:mark}

To explain the decrease in the equivalent width of the circumstellar gas absorption line during the photometric transit, consider a simple model which makes no strong assumptions about the geometry. It assumes that: 

(i) the area of the stellar disc has four different regions: that covered by both gas and dust, that covered only by gas, that covered only by dust and that with a clean line of sight; 

(ii) the stellar disc is assumed to be of uniform brightness (a simplification that ignores limb darkening)



(iii) the dust attenuates the brightness uniformly across the area it covers due to an optical depth $\tau_d$, and likewise for the gas with an optical depth $\tau_g$.

\begin{figure}
    \centering
    \includegraphics[width=1.05\columnwidth]{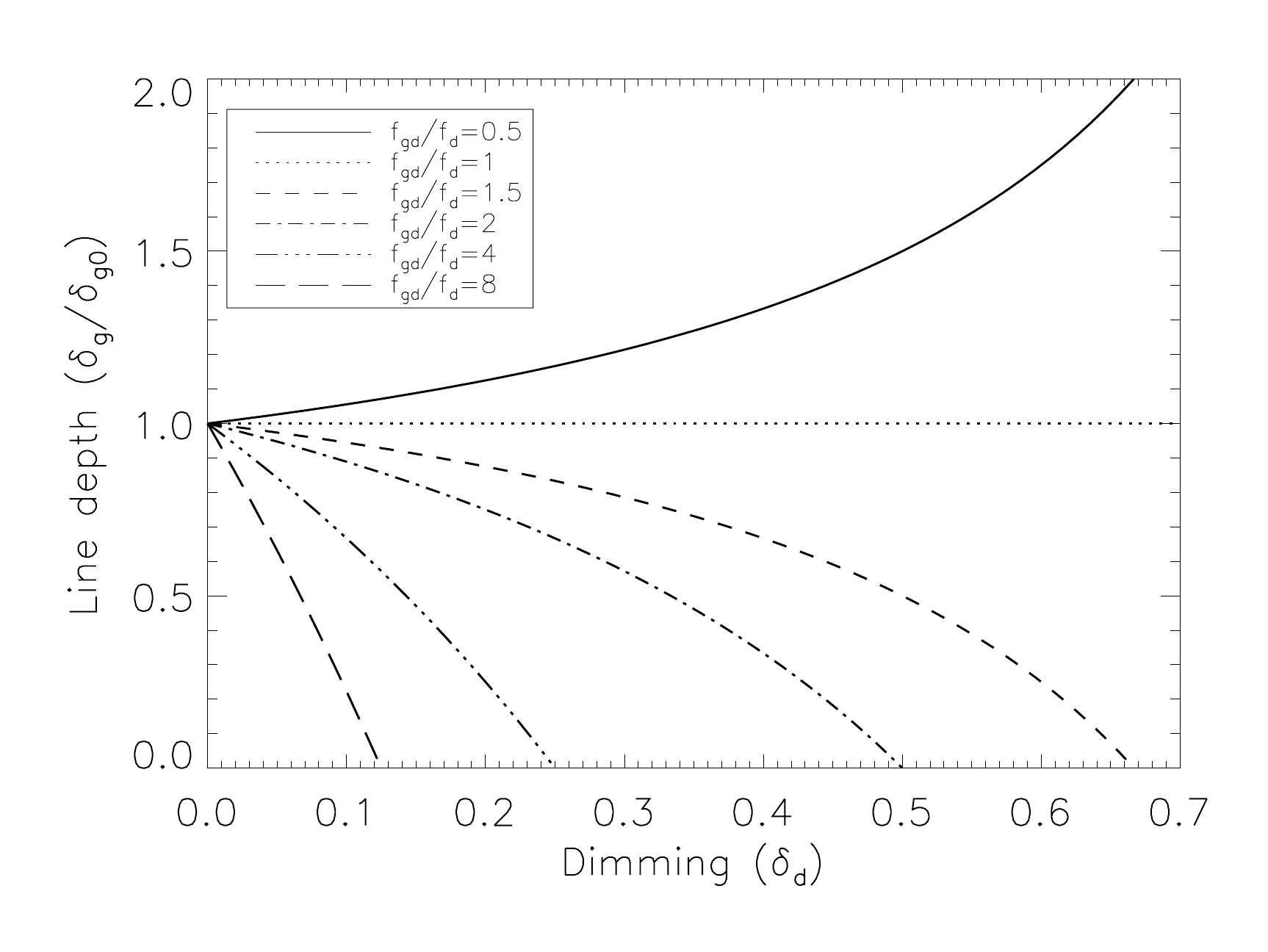}
    \caption{Fractional change in the line depth of a circumstellar gas absorption line $\delta_{\rm g}/\delta_{\rm g0}$ as a function of the (observable) level of dimming due to dust $\delta_{\rm d}$. This change depends only on $f_{\rm gd}/f_{\rm d}$, which is the ratio between the fraction of the gas causing the line that is covered by dust and the fraction of the star that is covered by dust.}
    \label{fig:dg}
\end{figure}

In this way, dust covering the fraction $f_\mathrm{d}$ of the stellar disc would cause an achromatic (see Sect.\,\ref{sec:grey}) dimming $\delta_{\rm d}=f_{\rm d}[1-\exp{(-\tau_{\rm d})}]$. Similarly, the circumstellar gas would absorb light, resulting in a line depth in the absence of dimming of $\delta_{\rm g0} = f_{\rm g}[1-\exp{(-\tau_{\rm g})}]$, where $f_{\rm g}$ is the fraction of the stellar disc covered by gas. Here we assume that the optical depth of the gas is sufficiently small to cause negligible attenuation in the continuum outside the line. With the above assumptions, the resulting line depth when the observed level of dimming is $\delta_{\rm d}$ is given by
\begin{equation}
  \delta_{\rm g} = \delta_{\rm g0} \left[ \frac{ 1-\delta_{\rm d}(f_{\rm gd}/f_{\rm d}) }{ 1-\delta_{\rm d} } \right], \label{eq:dg}
\end{equation}
where $f_{\rm gd}$ is the fraction of gas covered by dust. Therefore, the fractional change in the line depth is independent of the optical depths of the dust and the gas, it only depends on the observed level of dimming and the ratio $f_{\rm gd}/f_{\rm d}$. We plotted Eq.~\ref{eq:dg} in Fig.~\ref{fig:dg}, showing that during dimming:

-- there is no change in line depth if $f_{\rm gd}/f_{\rm d}=1$ (i.e. the dust covers equal fractional areas of gas and stellar surface)

 
-- the line depth increases if $f_{\rm gd}/f_{\rm d}<1$ (i.e. the dust covers more of stellar surface than what it covers of the gas)


-- the line depth decreases if $f_{\rm gd}/f_{\rm d}>1$ (i.e. the dust covers a greater fraction of gas than that corresponding to the stellar surface) 


Here we reported a significant decrease in the average equivalent width when the dimming reached  $50$\,per\,cent (B dip). This detection would constrain the ratio $f_{\rm gd}/f_{\rm d}$ between $1.5$ and $2$ (Fig.\,\ref{fig:dg}), meaning that the fraction of the gas covered by dust is around 1.5--2 times greater than the fraction of the star that is covered by dust. This implies that the geometry is such that the projected distribution of the gas and dust are correlated, and since the level of dimming reaches $50$\,per\,cent (and so $f_\mathrm{d}$ must be in the range $0.5$--$1$), the majority of the gas must be at locations where there is also dust when seen in projection against the star.


As we made no assumptions regarding the geometry of the system, the same change of $\delta_\mathrm{g}$ can be reached if the gas is being located closer to the white dwarf than the dust, or vice versa. If dust is closer, it would dim a fraction of the light before it reaches the gas. Alternatively, if the gas is closer to the white dwarf than the dust, the light affected by circumstellar absorption lines caused by the gas would afterwards be attenuated by the dust. 




\bsp	
\label{lastpage}
\end{document}